# Grain Size and Temperature-Dependent Response of 5-mol% Gd-Doped Ceria to Swift Heavy Ion Irradiation with 80 MeV Ag and 120 MeV Au Ions.

Waseem Ul Haq[a], V. Grover[b], Sanjay Kedia[c], Santanu Ghosh[a,**]

[a] Department of Physics, Indian Institute of Technology (IIT) Delhi, New Delhi 110016 India.

[b] Chemistry Division, Bhabha Atomic Research Centre (BARC), Mumbai 400085. India.

[c] Material Science Group, Inter-University Accelerator Centre (IUAC), New Delhi 110067, India.

## Abstract

The response of 5-mol% Gd-doped ceria to swift heavy ion beam irradiation has been studied to observe the effects of changes in ion energies and environmental temperature. The study involved irradiating two different grain sizes (nano and bulk) with two different ion energies: 80 MeV Ag and 120 MeV Au. Additionally, a comprehensive analysis of Gd-doped ceria's response to ion beam irradiation at high temperatures (1000 K) was conducted, taking into account the effect of grain size dependency. Based on GIXRD and Raman spectroscopy, it is evident that electronic excitation from 80 MeV Ag and 120 MeV Au ions at a fluence $1 \times 10^{14}$ ions/cm$^2$ caused damage to Gd-doped ceria samples. However, bulk grain size shows significant stability against SHI in all cases. These findings align with thermal spike simulations and indicate the formation of ion tracks due to electronic excitation by Swift Heavy Ion beam irradiation.

## Introduction

In the face of ever-increasing demand for nuclear energy, the hunt for materials that can endure the extreme radiation environments of the nuclear industry has become a vital challenge. When a high-energy ion beam with sufficient kinetic energy bombards a substance, the substance undergoes transformation through a variety of physical processes. Radiation-induced point defects (interstitials and vacancies) are produced during service [1] and can combine to form voids, stacking fault tetrahedra, and interstitial clusters [2–7]. This could ultimately contribute to the primary causes of material failure—swelling, hardening, amorphization, and embrittlement [8,9]. The material's thermal stability is also a significant concern when deciding on a material because nuclear reactors frequently operate beyond room temperature (typically ~ 1123K). Environmental temperature significantly influences defect migration, with irradiation-generated defects at elevated temperatures exhibiting a higher propensity to aggregate due to enhanced defect migration possibilities. This leads to material damage exceeding that at lower or ambient temperatures. Additionally, these defects recombine, causing annealing and reducing overall damage [10,11].

The initial reactivity control of a nuclear reactor is of paramount importance to ensure optimal performance and longevity in a nuclear power plant. Reactivity, which refers to the measure of the reactor's ability to sustain a chain reaction, can be effectively managed using burnable poisons. These poisons act as additional neutron absorbers when incorporated into fissile materials [12]. Among the various approaches to enhance reactivity control, several fuel utilization techniques have been developed, involving the inclusion of neutron absorbers along with fissile materials. Notably, Gadolinium, a fission product that can form solid solutions with uranium dioxide, has been employed as a dopant in $UO_2$ fuels for boiling water reactors [13,14]. The compound $Gd_2O_3$, composed of Gadolinium oxide, serves as the dopant. This choice is motivated by the large thermal absorption cross-section exhibited by Gadolinium isotopes, measuring 49,000 barns when considering the natural abundance of Gd isotopes [13,15]. This enhanced neutron absorption capability significantly contributes to the reactor's control and regulation of reactivity during its operational lifetime [13,16].

$CeO_2$, known for its fluorite crystal structure, has been proposed as a viable non-radioactive substitute for $PuO_2$ due to its exceptional resistance to radiation-induced amorphization [17,18]. Cerium ions exhibit two adaptable oxidation states with localized f electrons: $Ce^{3+}$ ($4f^1$) and $Ce^{4+}$ ($4f^0$), which closely resemble $Pu^{3+}$ ($5f^5$) and $Pu^{4+}$ ($5f^4$), respectively. The reduction of cerium ions and the consequent generation of oxygen vacancies in $CeO_2$ were assessed using x-ray photoelectron spectroscopy (XPS) [19]. The incorporation of Gadolinium oxide into cerium oxide has been extensively studied mainly as a solid oxide fuel cell application, ensuring high performance and continued usefulness in long operational life [20,21]. The phase studies [22], the chemical and physical properties of Gd-doped Ceria have been thoroughly investigated [23–25].

The large number of grain boundaries and the high surface-to-volume ratio of nanomaterials have made them a focus of research in recent years. Due to nanoparticles' enhanced electrical, optical, mechanical, and chemical capabilities, quite significant technological improvements have been made so far [26]. Ion beam irradiation is an established technique for modifying the characteristics of materials and researching radiation effects. The radiation stability of the material is primarily influenced by grain size and environmental temperature dependence [11]. It has been proposed that nanocrystalline materials will be more radiation-tolerant than their bulk (single-crystalline / micro-crystalline) counterparts owing to their relatively large volume-fraction of grain boundaries and "short" diffusion lengths. The grain boundaries serve as an effective "sink" for the defects (vacancies and interstitials) created during irradiation if the grain size is sufficiently comparable in diffusion length. Recent experimental and computational studies have shown enhanced radiation tolerance in nanomaterials. Advanced simulations show that interaction between GBs and irradiation-induced defects reduces radiation damage through a 'loading-unloading' mechanism. Interstitial defects are absorbed by GBs, which act as a source and annihilate vacancies, leading to healing and lowering radiation damage [27]. However, the critical point must be made is that in these publications[28–30], the irradiations were carried out with low energy ions, i.e., where the energy loss is predominantly nuclear energy loss ($S_n$) and collision cascades generate defects.

Yet another crucial point must be made: the irradiations carried out with high energy ions, i.e., where the electronic energy loss $S_e$ largely dominates, have exhibited a stark difference in contrast to the abovementioned results. Though nanocrystalline materials contain a more significant number of GBs, they were found to be less radiation-tolerant under the $S_e$ regime. In high-energy ion beam irradiation, a substance is bombarded with charged particles with sufficient kinetic energy (~ 10-100 MeV). These ion beams penetrate the material's surface far enough to cause atomic-scale structural modifications. The incident ion primarily interacts with the material's electronic sub-system prior to transferring its energy with the lattice sub-system. The lattice temperature increases as a result of the electron-phonon interaction, forming a cylindrical ion tract (~10 nm) inside the material's structure. The electron-phonon coupling constant is strictly affected by the grain size of the material. As the material is downsized, the phonon means the free path is more restricted via grain boundaries. Thus, leading to a more substantial thermal spike in nano-grained size material.

Another factor that governs heat dissipation after a temperature spike is thermal conductivity, which also changes with the variation in the grain size of the material. Thus, the rise, fall, and duration of thermal spike temperature, as well as irradiation-induced damage, depend on the electron-phonon coupling factor (g) and lattice thermal conductivity. Furthermore, it has been speculated that the crystallite/grain size and the surrounding temperature influence these factors. Therefore, understanding the physical mechanisms responsible for irradiation damage is a critical step toward improving material radiation tolerance.

The high-temperature environment in nuclear reactors necessitates the application of highly stable nuclear reactor material. Even though numerous studies have been devoted to the investigation of the radiation tolerance of Gd-doped ceria against Se-dominated irradiations. Tahara *et al.* have observed radiation damage at low concentrations against 200 MeV $Xe^{14+}$ ion irradiation[31]. In contrast, Costantini *et al.* have not observed any significant change except a decrease in the intensity of the $F_{2g}$ band observed by Raman Spectroscopy [30]. In another work, the behavior of GDC has been studied with change in the mol% of Gd- in Ceria only, where there is a phase transformation from F-type to C-type against 200 MeV $Xe^{14+}$ ion irradiation [18].

To the best of our knowledge, the radiation tolerance of nanocrystalline GDC against $S_e$ at nuclear reactor temperature/ high temperature with the dependence on grain size has not yet been thoroughly explored. To that purpose, polycrystalline Gd-doped ceria pellets of different grain sizes were bombarded with 80MeV $Ag^{6+}$ ions at ~1000 K to test their radiation tolerance against fission fragments. The irradiations were performed at the average nuclear reactor temperature (~1000 K) to better imitate a nuclear reactor environment.

**Experimental**

In the current research, we employed the gel-combustion technique to synthesize the cubic fluorite structure of 5 mol% Gadolinium doped Ceria (GDC) in the form of powders. Cerium oxide ($CeO_2$) and Gadolinium oxide ($Gd_2O_3$) were utilized as the oxidants, while glycine served as the fuel in this synthesis process. Initially, the required amounts of ($CeO_2$) and ($Gd_2O_3$) were dissolved in nitric acid, and subsequently, glycine was added to the solution. To achieve fuel-deficient combustion, the amount of glycine used was intentionally less than the stoichiometric requirement.

$$0.95 Ce(NO_3)_3 \cdot 6H_2O + 0.05\ Gd(NO_3)_3 \cdot 6H_2O + 1.6\ C_2H_5NO_2$$
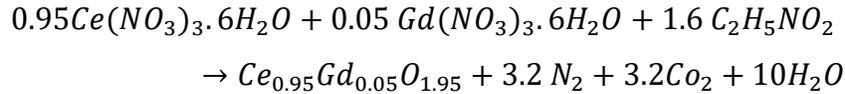
$$\to Ce_{0.95}Gd_{0.05}O_{1.95} + 3.2\ N_2 + 3.2 Co_2 + 10 H_2O$$

The GDC powder was calcined in ambient air/argon gas by placing the pellets in a quartz tube and heating at 600 °C. To achieve different particle/grain sizes and microstructures that can affect the response to irradiation, the GDC nanopowder was divided into two parts, formed into pellets, and subjected to heating at temperatures of 800 °C for 2 hours and 1300 °C for 8 hours. The samples annealed at 800 °C and 1300 °C would be denoted as S800 and S1300, respectively. Both small and large grain GDC samples were exposed to high-energy ion beam irradiation at room temperature and high temperature (~1000 K) using a pelletron facility at the Inter-University Accelerator Centre in New Delhi. The irradiation technique involved using 80 MeV Ag ions at a fluence of $1\times10^{14}$ ions/cm$^2$. This was done to simulate the fission fragment process that occurs in a nuclear reactor environment. We have calculated the values of electronic and nuclear energy loss using Stopping and Range of Ions in Matter (SRIM) software [32] for 5 mol% Gd-doped ceria, which has a density of 7.2 g/cm³. The calculated electronic energy loss for 80 MeV silver ions (Ag) is 18.7 keV/nm, while for 120 MeV gold ions (Au), it is 24.83 keV/nm. In comparison, the nuclear energy loss values are 1.5 keV/nm for silver and 4.6 keV/nm for gold. In typical nuclear reactor scenarios, the values of electronic energy loss (Se) and nuclear energy loss (Sn) are approximately 15.3 keV/nm and 16.9 keV/nm, respectively, when considering 70 MeV barium (Ba) and 100 MeV krypton (Kr) ions. Therefore, the selected ion beam energies are sufficient to induce damage within the material.

The surface morphology of GDC samples was studied using field emission scanning electron microscopy (FESEM). The glancing angle X-ray diffraction (on a PANalytical X'Pert Pro diffractometer with Cu $K_\alpha$ radiation) in the standard $2\theta$ *Geometry* method was employed to primarily examine the crystalline behaviour of GDC in both pre- and post-irradiation. Raman spectra of pristine and irradiated samples were also recorded on the surface of the samples using a Confocal Renishaw Raman microscope with a laser excitation wavelength of 532 nm.

**Characterization (Pristine):**

**FESEM:** The surface morphology of GDC virgin samples has been investigated using Field Emission Scanning Electron Microscopy (FESEM). Figure 1 (a) shows that the S800 samples have particles ranging from 30-50 nm, whereas Figure 1 (b) shows that the particle size in S1300 rose to 2-5 µm. S1300-grained size samples exhibit distinct particle sizes, possess high density, and are classified as bulk materials. The change in particle size in both samples is attributed to the heat treatment applied after nanopowder synthesis.

**X-ray diffraction:** X-ray diffraction is a non-destructive technique used to investigate a material's crystallographic structure, characterize crystallized materials, and determine their structures. It is commonly used in material science for identifying phases and understanding unit cell dimensions, crystallite size, and lattice strain. The X-ray diffraction (XRD) technique was used to analyze the crystalline structure of the pristine GDC samples. Figure 2 (a) clearly demonstrates that both GDC samples S800 and S1300 exhibit a cubic fluorite structure. The absence of any additional peaks in the sample can be attributed to the difference in sintering temperature. After carefully examining the full width at half maximum (FWHM) of the XRD peak (111), it is evident that S1300 has a lower FWHM compared to S800, as shown in the inset of Figure 1 (a). This suggests that there is an enhancement in the size of the crystallite as the sintering temperature increases. The Debye Scherrer formula yields an estimated crystallite size of around 30 nm for S800 and approximately 50 nm for S1300. These XRD results are in good match with the observation that was made by the FESEM.

**Raman Spectroscopy:** Raman spectroscopy is a sensitive, non-invasive analytical technique that can identify chemical compounds in material structures based on the Raman effect, first elucidated by C.V. Raman in 1928 [33-34]. It can identify both organic and inorganic entities and provide insights into a material crystalline structure [35]. Figure 3 shows the recorded spectra of Raman spectroscopy of 5 mol% Gd-doped Ceria. The asymmetric high intensity peak indexed at 462 cm$^{-2}$ belongs to the $F_{2g}$ characteristic band representing the cubic fluorite structure of Gd-doped Ceria [36]. The $F_{2g}$ symmetric band corresponds to the symmetrical stretching vibration of Ce–O$_8$ units and contains the force constants of Ce–O and O–O. The Raman spectra indicate a small difference in the full width at half maximum (FWHM) or peak broadening in S800 and S1300. This is again a result of the differences in the heat treatment after the synthesis process. These findings align well with the results obtained from the XRD and field emission scanning electron microscopy (FESEM).

**Characterization (post-irradiation HT):**

A comprehensive analysis of the crystallographic changes in post-irradiated GDC samples was conducted through glancing angle X-ray diffraction. This investigation aimed to extract the alterations following both room-temperature (~300 K) and high-temperature (~1000 K) irradiation. The irradiation on all samples was carried out at a reasonably high fluence of 1 × 10$^{14}$ ions/cm$^2$ with 80 MeV Ag ions. Based on the GIXRD spectra shown in Figure 4, it is evident that none of the samples were significantly amorphized, even at the highest chosen fluence. This clearly demonstrates that Gd-doped ceria microstructures exhibit exceptional stability against swift heavy ion beam irradiation, irrespective of environmental temperature.

The close examination of the GIXRD spectra reveals (insets of Figure 4. (a) and 4. (b)) that there is peak broadening in all the samples irradiated at room temperature (RT) and high temperature (HT). The peak broadening is more significant in the samples irradiated at room temperature (RT) than at high temperature (HT). This suggests that the GDC samples sustain more damage at RT, indicating that the long-range order of crystallinity has been compromised. In XRD analysis, we focused on observing the alteration post-irradiation by examining the change in the Full Width at Half Maximum (FWHM) of the most intense peak. For this reason, we have calculated the FWHM change of the (111) peak for each sample and summarized the findings in Table No.1 The calculation of irradiation damage was performed using the equation:

$$Damage = \frac{FWHM_{(111)-irradiated} - FWHM_{(111)-pristine}}{FWHM_{(111)-pristine}} \qquad (i)$$

| Sample Name | Temperature | Damage % |
|---|---|---|
| S800 | RT | 31 % |
| S800 | HT | 20 % |
| S1300 | RT | 15 % |
| S1300 | HT | 10 % |

Beyond the reduction in damage observed at high temperature (HT), as previously discussed, two additional significant findings were noted: (i) At room temperature (RT), the nano-crystalline samples (S800) experienced substantially more damage than the bulk-like sample (S1300), contrasting with earlier research on how crystallite size influences radiation tolerance under low-energy irradiation. (ii) At HT, damage remained lower for the S1300 sample compared to S800; however, the reduction in damage for the nano-crystalline samples was unexpectedly greater than that for the bulk-like sample.

These XRD results reveal a crucial and surprising dependency of radiation response on both the sample's microstructure and the irradiation temperature. This will be thoroughly examined in relation to the inelastic thermal spike model discussed later.

**Raman Spectroscopy**

We performed Raman spectroscopy measurements to analyse the changes in crystal structure of Gd-doped ceria in both pristine and irradiated conditions. Figure 5 shows that there is a single intense symmetric band centered at 462 cm$^{-1}$ due to the $F_{2g}$ mode of the Ce–O present in Gd-doped Ceria. From Raman spectra, it was found that there is a slight increase in FWHM of all samples post-irradiation at fluence of 1 x 10$^{14}$ ions/cm$^2$. However, the significant broadening is more observable in nano-grained samples compared to bulk samples, suggesting that smaller grain size samples are more damaged to damage from SHI at both room temperature and high temperature.

The reduced structural changes in bulk and decreased damage at HT compared to RT consistently align with prior GIXRD findings. Therefore, the bulk grain size sample exhibits exceptional stability within an electronic energy loss regime.

**Electronic energy loss dependency**

**GIXRD:**

To simulate the environment of a nuclear reactor and the resulting fission fragments from various ion sources, we have specifically chosen two ion species (80 MeV Ag and 120 MeV Au ions) to assess the structural stability of nano as well as bulk grain sized Gd-doped ceria. Figure 6 shows the GIXRD spectra of Pristine vs irradiated (120 MeV Au) samples (S800 & S1300). It is clearly evident that cubic fluorite structure is significantly retained in all grain sizes irrespective of ion species used for irradiation. This has been inferred from the fact there is no evolution or disappearance of diffraction peaks due to electronic excitation due to 120 MeV Au ions at fluence $1 \times 10^{14}$ ions/cm$^2$. However, upon a close investigation of FWHM & peak asymmetry from GIXRD spectra, there is a change in the peak width and peak shift post-irradiation as shown in Figure 6 (b). These alterations are linked with the generation of lattice strain and therefore deterioration of crystallinity with high-energy ion beam irradiation. The impact of changes is particularly pronounced in samples with smaller grain sizes, regardless of ion doses.

The comparison of the figures 7 (a and b) clearly shows that the 120 MeV Au, with its higher energy, has led to significant damage in GDC samples of both grain sizes (nano and bulk) as compared to 80 MeV Ag. In the S800 sample, the noticeable decrease in intensity and peak asymmetry strongly suggests that the material underwent deformation due to high Se. However, fewer changes are observed with the irradiation of 80 MeV Ag, as shown in Figure 7 (a). This observation can be attributed to the direct relationship between the concentration of $S_e$ and the extent of damage incurred by the material. As depicted in Figure 7(b), it is apparent that the bulk-grained size sample (S1300) underwent some level of modification, albeit not as extensively as the nano-grained size sample (S800). A closer examination of the GIRDX peaks indicates that the sample irradiated with 120 MeV Au suffered more damage compared to the one irradiated with 80 MeV Ag. Upon high-energy ion beam irradiation, it has been observed that there is a direct correlation between energy levels and damage, as well as between grain size and damage. Specifically, higher energy levels result in increased damage, while smaller grain sizes lead to higher damage. Conversely, larger grain sizes are associated with decreased damage. These observations are thoroughly explained based on the thermal spike model in the subsequent section.

**Raman Spectroscopy**

Raman spectroscopy is a type of molecular spectroscopy that infers information from the materials by utilizing an incident visible laser. We have again performed Raman Spectroscopy measurements on pristine as well as irradiated 5-mol% Gd-doped Ceria. The Raman shifts at 462 cm$^{-1}$, attributable to the $F_{2g}$ mode, serve as compelling evidence of the presence of GDC samples, as clearly illustrated in Figure 8. The samples irradiated with 80 MeV Ag show stronger stability as there is a minor change in FWHM post-irradiation in both grained size samples. All these results are again consistent with GIXRD measurements. The possible reason behind such damage is explained based on the inelastic thermal spike model in the later section.

**In-elastic thermal spike Simulations**

When an ion interacts with a material, it significantly imparts its energy into the material, thereby modifying it based on the energy it carries. This transformative process underscores the profound impact ions can have on materials, making their interaction a vital area of study and exploration. The energy loss due to swift heavy ion beam irradiation has been explained based on inelastic thermal spike model.

The thermal spike model is a theoretical framework that explains energy transfer and temperature changes in materials exposed to high-energy radiation, such as ion or neutron irradiation. When energetic particles penetrate into a solid, they transfer energy along their paths, causing rapid localized hot zones and the creation of transitory thermal spikes. Thermal spikes can induce localized melting, amorphization, or other structural alterations in the material, depending upon the energy and duration of the spike. The model offers a framework for comprehending energy dissipation via electron-phonon interactions and heat conduction on extremely short durations, generally in the range of picoseconds. It is essential for estimating the radiation-induced effects in materials, including damage and defect generation, and phase changes, hence functioning as a vital instrument in materials science and radiation damage research.

The temperature fluctuations of the electrons and the lattice can be explained using two linked differential equations in cylindrical geometry [37]

$$\rho C_e(T_e)\frac{\partial T_e}{\partial t} = \nabla(K_e(T_e)\nabla T_e) - g(T_e - T) + A(r,t) \qquad (ii)$$

$$\rho C(T)\frac{\partial T}{\partial t} = \nabla(K(T)\nabla T) + g(T_e - T) \qquad (iii)$$

Specific heats and thermal conductivities of the electronic and lattice sub-system are represented by Ce, C, Ke, and K, respectively. Te and T stand for the electronic and lattice temperatures, while ρ represents the mass density. "g" denotes the electron-phonon interaction, and r and t are the distance from the center of the ion path and time, respectively. A(r,t) is the energy density deposition on electronic subsystem $\iint 2\pi r A(r,t)drdt = S_e$. In this research, we have performed thermal spike simulation to mimic the results obtained from experiments and characterization.

In general, the thermal spike model explains that energy dissipation in irradiated materials is significantly affected by two factors the strength of electron-phonon coupling and thermal conductivity.

**i). Electron-phonon coupling (g):** The intensity of electron-phonon coupling determines the rate of energy transfer from excited electrons to the lattice. When a high-energy ion traverses a material, it excites electrons to higher energy levels, resulting in a non-equilibrium condition. The intensity of electron-phonon coupling indicates the rate at which heated electrons transfer energy to lattice atoms, resulting in localized heating. Intense electron-phonon coupling facilitates swift energy transfer, resulting in a rapid increase in lattice temperature. This may increase the probability of localized melting, phase transitions, or defect generation. In contrast, poor coupling leads to diminished energy transfer, resulting in a more gradual temperature rise.

The efficiency of the energy transfer from the electrons to the lattice is determined by the correlation between the electron-phonon coupling constant, which is governed by an equation [38]

$$g = \frac{9 n_e k_b^2 T_D^2 V_f}{16 \lambda T E_f} \quad (iv)$$

where, $n_e$ stands for the electron number density, $k_b$ for Boltzmann's constant, $E_f$ for Fermi energy, T for ambient temperature, $T_d$ for Debye temperature, $V_f$ for Fermi velocity, lambda isthe mean diffusion length of excited electrons or for electron-phonon mean free path.

The electron-phonon coupling strength (g), which influences the lattice temperature increase, is determined by the electron-phonon mean free path (λ), or the average diffusion length of the electrons. We conducted i-TSM simulations for both S800 and S1300 grain sizes, as the electron-phonon coupling varies with different grain sizes.

It is well established that grain boundaries act as obstacles to electron movement. Consequently, the grain or crystallite size significantly affects the electron mean diffusion length (λ). As the grain size decreases, the number of grain boundaries increases, which leads to a considerable reduction in the electron-phonon mean free path. This relationship between crystallite size and (λ) is calculated using the formula below,

$$\lambda = \frac{\alpha d (1 - R)}{R} \quad (v)$$

Where, α is the angle between velocity vector of the electron and the material plane and R is the reflection coefficient of electrons striking the grain boundaries. We have estimated the value of lambda ($\lambda$) ~ 1.5 nm and ~ 4 nm (approx.) for S800 & S1300, respectively as calculated earlier [11,39].

**ii). Thermal conductivity** is also essential in influencing the parameter of the thermal spike. It affects the rate of heat conduction away from the area of localized heating. Materials with high thermal conductivity rapidly remove heat, resulting in a swift decrease of the thermal spike and constraining the impacted area. This can decrease the risk of substantial structural alterations or damage. Conversely, materials with less thermal conductivity preserve heat for a protracted duration, enabling the thermal spike to endure and potentially resulting in more significant consequences, such as prolonged melting or defect aggregation. The interaction between electron-phonon coupling strength and thermal conductivity influences the temperature distribution, duration, and spatial extent of thermal spikes, directly affecting the material's reaction to irradiation.

The equation that governs the effective thermal conductivity with the variation of grain size is as follows [40]

$$K = \frac{\frac{K_O}{1+\frac{\Lambda_O}{d^{.075}}}}{1+\frac{R_k}{d}\left[\frac{K_O}{1+\frac{\Lambda_O}{d^{.075}}}\right]} \qquad (vi)$$

Where, $R_k$ is the Kapitza thermal resistance, $\Lambda_O$ is the single crystal phonon mean free path, and $K_O$ is the single crystalline thermal conductivity. At room temperature, it has been noted that the thermal conductivity is more dominant in smaller grain size samples due to the variation of, $\Lambda_O$ is the single crystal phonon mean free path and is governed by the equation[11].

$$\Lambda_0 = \frac{20T_m a}{\gamma^2 T} \qquad (vii)$$

Where T is the irradiation temperature, $\gamma$ is the Gruneisen Constant, $T_m$ is the melting temperature. In equation (vii), it's clear that the irradiation temperature significantly influences the maintenance of thermal conductivity during irradiation. This underscores the crucial role of temperature control in optimizing thermal conductivity. Hence, we carried out thermal spike simulations to simulate ion beam irradiation at both room temperature (T=300 K) and at a high temperature of around (T=1000 K).

The necessary parameters for the simulation of the Thermal Spike Model (TSM) in Gd-doped ceria (5 mol% Gd) are currently unavailable in the literature. However, the parameters required for the inelastic Thermal Spike Model (i-TSM) for pure ceria ($CeO_2$) have been well-documented. Consequently, we conducted Thermal Spike Simulations exclusively for pure ceria. In these simulations, we assumed that the structural properties of pure ceria are analogous to those of Gd-doped ceria (5 mol% Gd), enabling us to investigate the grain size effects as previously described.

Additionally, we also calculated the theoretical lattice parameter from Vegard's Law by the formula[11],

$$a_{Vegard} = \frac{4}{\sqrt{3}[x_{Gd}r_{Gd} + (1-x_{Gd})r_{Ce} + (1-.25x_{Gd})r_0 + .25r_{Gd}r_{V_0}] + .9971} \quad (viii)$$

where $r_{Gd}$=0.1053nm, $r_{Ce}$=0.0970nm, $r_0$=0.1380nm, and $r_{V_0}$=0.1164nm are given as the ideal ionic radius of $Gd^{3+}$ ($r_{Gd}$), $Ce^{4+}$ ($r_{Ce}$), $O_2$ ($r_0$), and $V_O$, respectively. The estimated value of the theoretical lattice parameter from Eqn (viii) is .541 plus or minus .0001, which is again more or less consistent with the values obtained from Pristine S800 and S1300.

Meftah *et.al* [25] reported that electronic thermal conductivity and electronic specific heat are assumed as (2 J cm K) and (1 J/ cm K), respectively for insulators. The lattice-specific heat and thermal conductivity were taken from Andrew T et al.[26].

The lattice thermal conductivities used for simulations were obtained from [26] and input parameters for simulations in $CeO_2$ are detailed in table 1.

| Solid Density | 7.2 g/cm³ |
|---|---|
| **Melting temp.** | 2670 K |
| **Boiling temp.** | 3770 K |
| **Latent heat of fusion** | 470 j/g |

Table 1. Input parameters involved during thermal spike simulations.

Using heat diffusion equations, the temperature profile *T(r,t)* can be determined as a function of radial distance (*r*) and time (*t*). The track radius is often defined as the distance from the center where the temperature falls to a threshold value below which no permanent damage occurs. This threshold temperature can be related to the material's melting or damage temperature, and solving for this distance gives an estimate of the track radius.

**Results of i-TSM.**

I) Interplay of grain size and environmental temperature. Figure 9 & 10 depicts the outcomes of simulations that explore the impact of grain size and environmental temperatures i.e. at room temperature and high temperature, respectively. The samples irradiated at room temperature are labeled as GDC-S for Nano size and GDC-L for bulk size. It is evident that the sample bombarded at room temperature shows a significantly larger track radius compared to the sample irradiated under high environmental temperatures. Detailed results for various temperatures can be found in Table No. 02.

| Sample name | Track Radius | Thermal spike duration (ps) | Highest temperature(K) around 0 nm |
|---|---|---|---|
| **S800 RT** (GDC-S) | 3 nm | 9.998 | 11209 |
| **S800 HT** | 2.4 nm | 7.293 | 11780 |
| **S1300 RT** (GDC-L) | 1.6 nm | 5.995 | 3929 |
| **S1300 HT** | 1.4 nm | 5.532 | 4500 |

II) **Electronic energy loss dependency:** The results depicted in Figure No. 11 illustrate the thermal spike simulation outcomes for S800 and S1300 when subjected to two distinct ion energies of 120 MeV Au ions. It is observed that the track radius in S800 is notably larger than that in S1300 for both ion energies. Moreover, the reduced track radius in S1300 indicates that larger grain size is associated with decreased damage. Furthermore, the data suggests that higher ion energy induces greater damage compared to 80 MeV Ag ion, likely owing to the larger Se. The observed values of track radius and lattice temperature are summarized in Table No. 3.

| Sample name | Energy (MeV) | Thermal spike duration (ps) | Track Radius (nm) | Highest temperature(K) around 0 nm |
| --- | --- | --- | --- | --- |
| GDC-S | 80 | 9.998 | 3 | 11209 |
| S800 | 120 | 12.89 | 5 | 18000 |
| GDC-L | 80 | 5.995 | 1.6 | 3929 |
| S1300 | 120 | 8.75 | 4 | 12000 |

In the subsequent section, a comprehensive explanation of the Se dependency and the impact of environmental temperature will be provided. This will include how environmental temperature can affect, Se dependency and the mechanisms by which this interaction occurs.

**Discussion.**

The experimental results obtained from GIXRD and Raman Spectroscopy show a strong correlation with the in-elastic thermal spike simulations, demonstrating the excellent agreement between the two. The GIXRD (Grazing-Incidence X-ray Diffraction) spectra analysis and Raman Spectroscopic analysis of Gd-doped ceria indicate a substantial reduction in crystallinity in smaller grain sizes following irradiation. This effect is observed irrespective of variations in environmental temperatures and beam energies. The findings suggest that samples with a bulk/micro grain size sample exhibit more excellent stability when exposed to high-energy ions. These results align with those obtained from the inelastic thermal spike model. The observed phenomenon can be attributed to the influence of grain boundaries on the interaction between electrons and phonons, which in turn affects the material's thermal conductivity. When the grain size of a material is reduced, it leads to the formation of a larger number of grain boundaries within the material's structure. These grain boundaries then serve as sites for electron scattering, affecting the material's electron-phonon mean free path ($\lambda$). The relationship expressed in equation (v) shows that the parameter lambda (λ) is directly influenced by the grain size (d). As a result, in smaller grain sizes, the e-ph parameter (g) exhibits increased strength, which in turn leads to the formation of a higher lattice temperature. This demonstrates the intricate interplay between grain size and lattice temperature within the context of ion beam irradiation.

At higher environmental temperatures, such as those experienced in high-temperature environments, the samples exhibited a reduced level of damage in comparison to their behavior at room temperature. This phenomenon can be attributed to the presence of smaller grain sizes, which possess a larger fraction of grain boundaries. These grain boundaries serve as an obstacle to the mean free path of phonons as indicated by equation no. (viii). As a result, the existence of numerous grain boundaries leads to a reduction in thermal conductivity, consequently resulting in a decrease in the magnitude of the thermal spike. This behavior indicates the complex interplay between grain size, grain boundary density, and thermal transport properties in materials exposed to varying environmental temperatures as represented schematically in Figure 11. The diagram in the figure illustrates how ion beams interact and how the phonon mean free path varies at both room temperature (RT) and high temperature (HT). This indicates that irradiation damage is surprisingly low at high temperatures, suggesting that Gd-doped Ceria exhibits better stability under irradiation at elevated temperatures.

**Conclusion.**

To summarise, the GIXRD and Raman Spectroscopy experimental results of 5 mol% Gd-doped ceria irradiated at swift heavy ion beam irradiation (80 & 120 MeV Au ions) show a strong correlation with in-elastic thermal spike simulations. The results suggest that samples with a bulk/micro grain size exhibit better stability when exposed to high-energy ions as compared to nano grain size samples. This is due to the influence of grain boundaries on the interaction between electrons and phonons, which affects the material's thermal conductivity. Smaller grain sizes (S800) lead to more grain boundaries, which serve as electron scattering sites and affect the material's electron-phonon mean free path. At higher environmental temperatures (1000 K), the samples show reduced damage due to the presence of smaller grain sizes, which indicates that bulk grain size sample has better radiation stability against 80 MeV Ag ion irradiation.

**Acknowledgments**

We express profound gratitude to all members of the Material Science Group at the Inter-University Accelerator Centre (IUAC) for their invaluable support throughout the High-energy ion beam irradiation experiments. Waseem Ul Haq extends sincere gratitude to the University Grants Commission (UGC) India for the assistance of fellowships (JRF-SRF) during the course of the research. Our sincere thanks is also extended to the characterization facilities utilized in this research, namely X-ray Diffraction (XRD) from the Department of Physics, Raman Spectroscopy under the FIST UFO Scheme, and Field-Emission Scanning Electron Microscopy (FE-SEM) from the Central Research Facility (CRF) at the Indian Institute of Technology Delhi (IITD).

## References

[1] G.D. Watkins, EPR Observation of Close Frenkel Pairs in Irradiated ZnSe, Phys. Rev. Lett. 33 (1974) 223–225. https://doi.org/10.1103/PhysRevLett.33.223.

[2] Observation of the One-Dimensional Diffusion of Nanometer-Sized Dislocation Loops | Science, (n.d.). https://www.science.org/doi/full/10.1126/science.1145386?casa_token=-LjPWUukGkgAAAAA%3AEIv6zUBW2f5pv_AY8QuZ6gCp0EhbqgCjzs_f0JAgHTb6fi_SIsmtu65z-bKqZGT_epmE4y5H_XmiDjqc (accessed July 25, 2023).

[3] How Does Radiation Damage Materials? | Science, (n.d.). https://www.science.org/doi/full/10.1126/science.1150394?casa_token=7jFfcApDQ-cAAAAA:uqOuBx6r-1IJKs_o0VS1grV5HvGmRv5-314jWr8zE_5iAydofb4ey0FCJ74nUiz_caq1VjrcdS4sQ5mY (accessed July 25, 2023).

[4] M. Victoria, N. Baluc, C. Bailat, Y. Dai, M.I. Luppo, R. Schӓublin, B.N. Singh, The microstructure and associated tensile properties of irradiated fcc and bcc metals, J. Nucl. Mater. 276 (2000) 114–122. https://doi.org/10.1016/S0022-3115(99)00203-2.

[5] One-Dimensional Fast Migration of Vacancy Clusters in Metals | Science, (n.d.). https://www.science.org/doi/full/10.1126/science.1148336?casa_token=XmaZKUVQr8EAAAAA%3Aca1on8ixCrg3cs1L5hjQNSFOZ_uDI18Rer7OdROlD4sc3PAMFvszMTys_XSPQamqvwX6Bdqlo7ctcr3a (accessed July 25, 2023).

[6] T. Diaz de la Rubia, H.M. Zbib, T.A. Khraishi, B.D. Wirth, M. Victoria, M.J. Caturla, Multiscale modelling of plastic flow localization in irradiated materials, Nature 406 (2000) 871–874. https://doi.org/10.1038/35022544.

[7] One-dimensional atomic transport by clusters of self-interstitial atoms in iron and copper: Philosophical Magazine: Vol 83, No 1, (n.d.). https://www.tandfonline.com/doi/abs/10.1080/01418610210000016793?casa_token=NHcrMjSpG6UAAAAA:JYceuYj2rL7Apbxxs1qs-O159Af3v_dIuMkha4bCdrWer4LhpZ_LwJnqXjTBb3rNqPHYq5fMKV6gf_s (accessed July 25, 2023).

[8] G.R. Odette, M.J. Alinger, B.D. Wirth, Recent Developments in Irradiation-Resistant Steels, Annu. Rev. Mater. Res. 38 (2008) 471–503. https://doi.org/10.1146/annurev.matsci.38.060407.130315.

[9] K.E. Sickafus, R.W. Grimes, J.A. Valdez, A. Cleave, M. Tang, M. Ishimaru, S.M. Corish, C.R. Stanek, B.P. Uberuaga, Radiation-induced amorphization resistance and radiation tolerance in structurally related oxides, Nat. Mater. 6 (2007) 217–223. https://doi.org/10.1038/nmat1842.

[10] P. Kalita, Grain size and environmental temperature effects on the consequence of ion energy loss in materials a case study of cubic zirconica, Indian Institute of Technology Delhi, 2020. https://libcat.iitd.ac.in/cgi-bin/koha/opac-detail.pl?biblionumber=167514.

[11] P. Kalita, S. Ghosh, U.B. Singh, P.K. Kulriya, V. Grover, R. Shukla, A.K. Tyagi, G. Sattonnay, D.K. Avasthi, Investigating the effect of material microstructure and irradiation temperature on the radiation tolerance of yttria stabilized zirconia against high energy heavy ions, J. Appl. Phys. 125 (2019) 115902. https://doi.org/10.1063/1.5080934.

[12] E. Hellstrand, Burnable poison reactivity control and other techniques to increase fuel burnup in LWR fuel cycles, Trans Am Nucl Soc U. S. 40 (1982). https://www.osti.gov/biblio/6948453 (accessed January 2, 2023).

[13] S.F. Mughabghab, Thermal neutron capture cross sections resonance integrals and g-factors, (2003). https://www.osti.gov/etdeweb/biblio/20332542 (accessed January 2, 2023).

[14] F.B. Skogen, R.G. Grummer, L.A. Nielsen, Operating experience with Exxon nuclear supplied gadolinia in pressurized water reactors, Trans. Am. Nucl. Soc. (1982) 194–198.

[15] W. Böhm, H.-D. Kiehlmann, A. Neufert, M. Peehs, $Gd_2O_3$ up to 9 weight percent, an established burnable poison for advanced fuel management in pressurized water reactors, Kerntechnik 50 (1987) 234–240. https://doi.org/10.1515/kern-1987-500411.

[16] S.K. Sharma, V. Grover, A.K. Tyagi, D.K. Avasthi, U.B. Singh, P.K. Kulriya, Probing the temperature effects in the radiation stability of $Nd_2Zr_2O_7$ pyrochlore under swift ion irradiation, Materialia 6 (2019) 100317. https://doi.org/10.1016/j.mtla.2019.100317.

[17] Comprehensive Nuclear Materials, Elsevier, 2020.

[18] P. Seo, K. Yasuda, S. Matsumura, N. Ishikawa, G. Gutierrez, J.-M. Costantini, Microstructure evolution in 200-MeV Xe ion irradiated $CeO_2$ doped with $Gd_2O_3$, J. Appl. Phys. 132 (2022) 235902. https://doi.org/10.1063/5.0121951.

[19] Role of Microstructure and Surface Defects on the Dissolution Kinetics of $CeO_2$, a $UO_2$ Fuel Analogue | ACS Applied Materials & Interfaces, (n.d.). https://pubs.acs.org/doi/full/10.1021/acsami.5b11323 (accessed July 24, 2023).

[20] H. Inaba, H. Tagawa, Ceria-based solid electrolytes, Solid State Ion. 83 (1996) 1–16. https://doi.org/10.1016/0167-2738(95)00229-4.

[21] M. Mogensen, N.M. Sammes, G.A. Tompsett, Physical, chemical and electrochemical properties of pure and doped ceria, Solid State Ion. 129 (2000) 63–94. https://doi.org/10.1016/S0167-2738(99)00318-5.

[22] V. Grover, A.K. Tyagi, Phase relations, lattice thermal expansion in $CeO_2$–$Gd_2O_3$ system, and stabilization of cubic gadolinia, Mater. Res. Bull. 39 (2004) 859–866. https://doi.org/10.1016/j.materresbull.2004.01.007.

[23] A. Banerji, V. Grover, V. Sathe, S.K. Deb, A.K. Tyagi, $CeO_2$–$Gd_2O_3$ system: Unraveling of microscopic features by Raman spectroscopy, Solid State Commun. 149 (2009) 1689–1692. https://doi.org/10.1016/j.ssc.2009.06.045.

[24] A. Kossoy, Q. Wang, R. Korobko, V. Grover, Y. Feldman, E. Wachtel, A.K. Tyagi, A.I. Frenkel, I. Lubomirsky, Evolution of the local structure at the phase transition in $CeO_2$-$Gd_2O_3$ solid solutions, Phys. Rev. B 87 (2013) 054101. https://doi.org/10.1103/PhysRevB.87.054101.

[25] V. Grover, A.K. Tyagi, Sub-solidus phase equilibria in the $CeO_2$–$ThO_2$–$ZrO_2$ system, J. Nucl. Mater. 305 (2002) 83–89. https://doi.org/10.1016/S0022-3115(02)01157-1.

[26] M. Singh, W.U. Haq, S. Bishnoi, B.P. Singh, S. Arya, A. Khosla, V. Gupta, Investigating photoluminescence properties of $Eu^{3+}$ doped $CaWO_4$ nanoparticles via $Bi^{3+}$ amalgamation for w-LEDs application, Mater. Technol. 37 (2022) 1051–1061. https://doi.org/10.1080/10667857.2021.1918866.

[27] X.-M. Bai, A.F. Voter, R.G. Hoagland, M. Nastasi, B.P. Uberuaga, Efficient Annealing of Radiation Damage Near Grain Boundaries via Interstitial Emission, Science (2010). https://doi.org/10.1126/science.1183723.

[28] Radiation Tolerance of Nanocrystalline Ceramics: Insights from Yttria Stabilized Zirconia | Scientific Reports, (n.d.). https://www.nature.com/articles/srep07746 (accessed July 24, 2023).

[29] T.D. Shen, S. Feng, M. Tang, J.A. Valdez, Y. Wang, K.E. Sickafus, Enhanced radiation tolerance in nanocrystalline $MgGa_2O_4$, Appl. Phys. Lett. 90 (2007) 263115. https://doi.org/10.1063/1.2753098.

[30] M. Jin, P. Cao, S. Yip, M.P. Short, Radiation damage reduction by grain-boundary biased defect migration in nanocrystalline Cu, Acta Mater. 155 (2018) 410–417. https://doi.org/10.1016/j.actamat.2018.05.071.

[31] Y. Tahara, B. Zhu, S. Kosugi, N. Ishikawa, Y. Okamoto, F. Hori, T. Matsui, A. Iwase, Study on effects of swift heavy ion irradiation on the crystal structure in $CeO_2$ doped with $Gd_2O_3$, Nucl. Instrum. Methods Phys. Res. Sect. B Beam Interact. Mater. At. 269 (2011) 886–889. https://doi.org/10.1016/j.nimb.2010.12.032.

[32] J.F. Ziegler, M.D. Ziegler, J.P. Biersack, SRIM – The stopping and range of ions in matter (2010), Nucl. Instrum. Methods Phys. Res. Sect. B Beam Interact. Mater. At. 268 (2010) 1818–1823. https://doi.org/10.1016/j.nimb.2010.02.091.

[33] R.C. Maher, L.F. Cohen, P. Lohsoontorn, D.J.L. Brett, N.P. Brandon, Raman Spectroscopy as a Probe of Temperature and Oxidation State for Gadolinium-Doped Ceria Used in Solid Oxide Fuel Cells, J. Phys. Chem. A 112 (2008) 1497–1501. https://doi.org/10.1021/jp076361j.

[34] C.V. Raman, K.S. Krishnan, A New Type of Secondary Radiation, Nature 121 (1928) 501–502. https://doi.org/10.1038/121501c0.

[35] L.A.K. Staveley, The Characterization of Chemical Purity: Organic Compounds, Elsevier, 2016.

[36] M. Singh, D. Zappa, E. Comini, NiO-GDC nanowire anodes for SOFCs: novel growth, characterization and cell performance, Mater. Adv. 3 (2022) 5922–5929. https://doi.org/10.1039/D2MA00317A.

[37] P. Kalita, S. Ghosh, G. Sattonnay, U.B. Singh, V. Grover, R. Shukla, S. Amirthapandian, R. Meena, A.K. Tyagi, D.K. Avasthi, Role of temperature in the radiation stability of yttria stabilized zirconia under swift heavy ion irradiation: A study from the perspective of nuclear reactor applications, J. Appl. Phys. 122 (2017) 025902. https://doi.org/10.1063/1.4993177.

[38] Large electronic sputtering yield of nanodimensional Au thin films: Dominant role of thermal conductivity and electron phonon coupling factor: Journal of Applied Physics: Vol 121, No 9, (n.d.). https://aip.scitation.org/doi/full/10.1063/1.4977845?casa_token=Y8f6oPuCiX0AAAAA%3AAOBleKYTeE-vdNTK0Dzr9UOjXtYaOZJ8_qFHX_PdmzMJEK90e9orfOh0LWYBEzqYc8_rbY0BgFXx (accessed November 22, 2022).

[39] W.U. Haq, V. Grover, P. Kalita, R. Shukla, F. Singh, S.K. Srivastava, S. Shukla, S. Ghosh, Study on damage of Gd2O3–CeO2 under electronic energy loss: comparison between bulk-like and nanostructure, Phys. Chem. Chem. Phys. 26 (2024) 5311–5322. https://doi.org/10.1039/D3CP05252D.

[40] H. Dong, B. Wen, R. Melnik, Relative importance of grain boundaries and size effects in thermal conductivity of nanocrystalline materials, Sci. Rep. 4 (2014) 7037. https://doi.org/10.1038/srep07037.

# Supporting figures

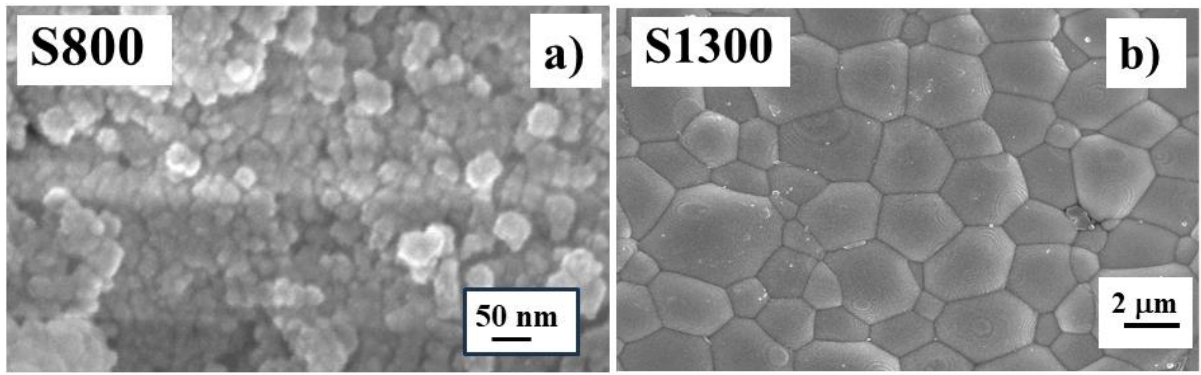

**Figure 1.** *FESEM images of 5 mol% Gd-doped ceria. a). S800- Nano-grained size sample, and b). S1300 bulk grain size sample.*

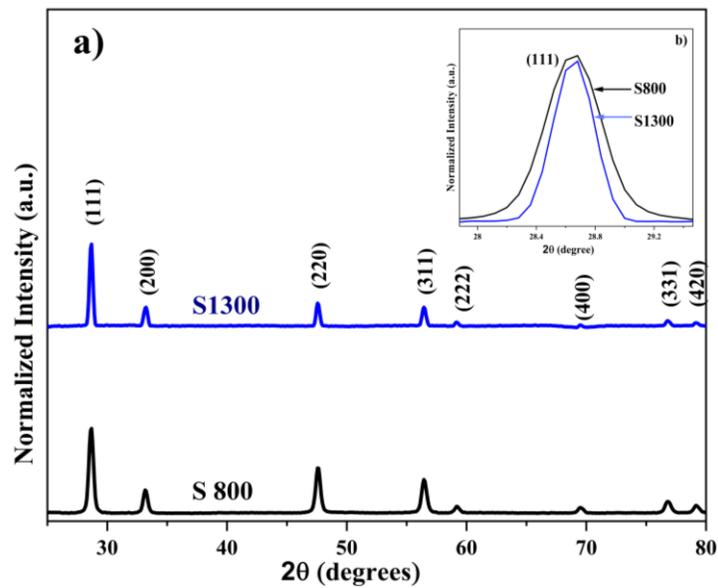

**Figure 2.** (a) GIXRD pattern of pristine 5 mol% Gd- doped Ceria GDC-S and GDC-L; (b) Zoomed view of the (111) diffraction maxima for both samples.

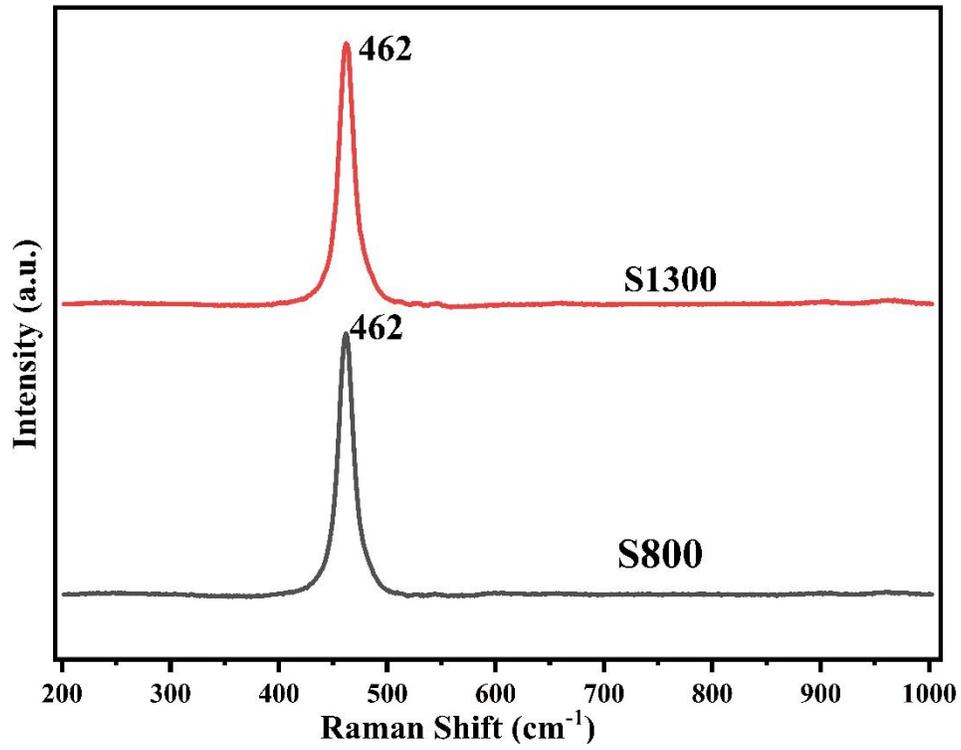

**Figure 3.** Recorded spectra of Raman spectroscopy of 5 mol% Gd-doped Ceria pristine S800 and S1300.

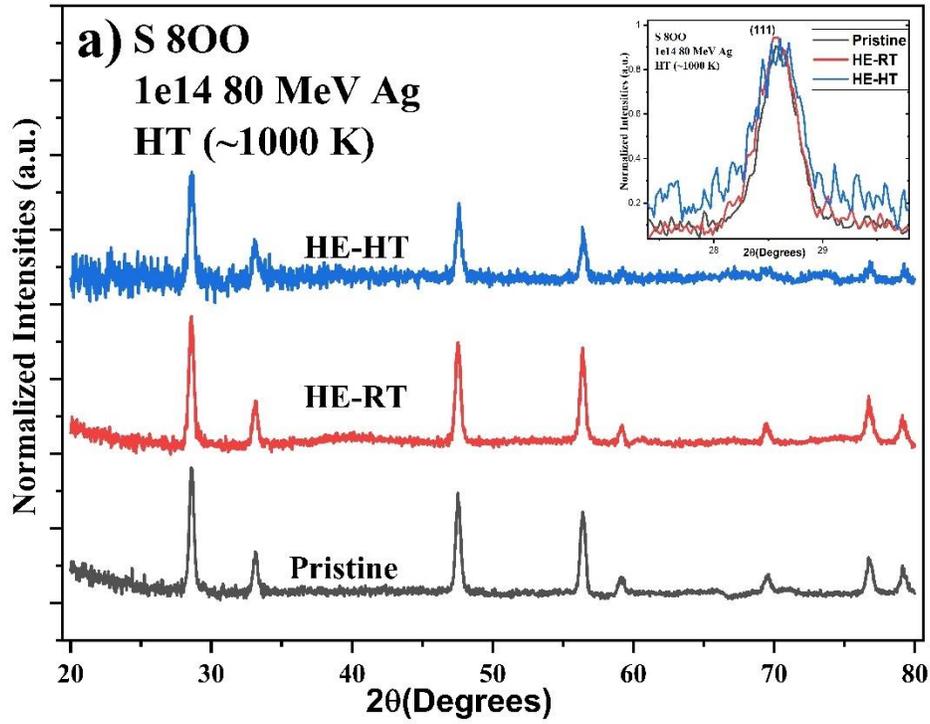
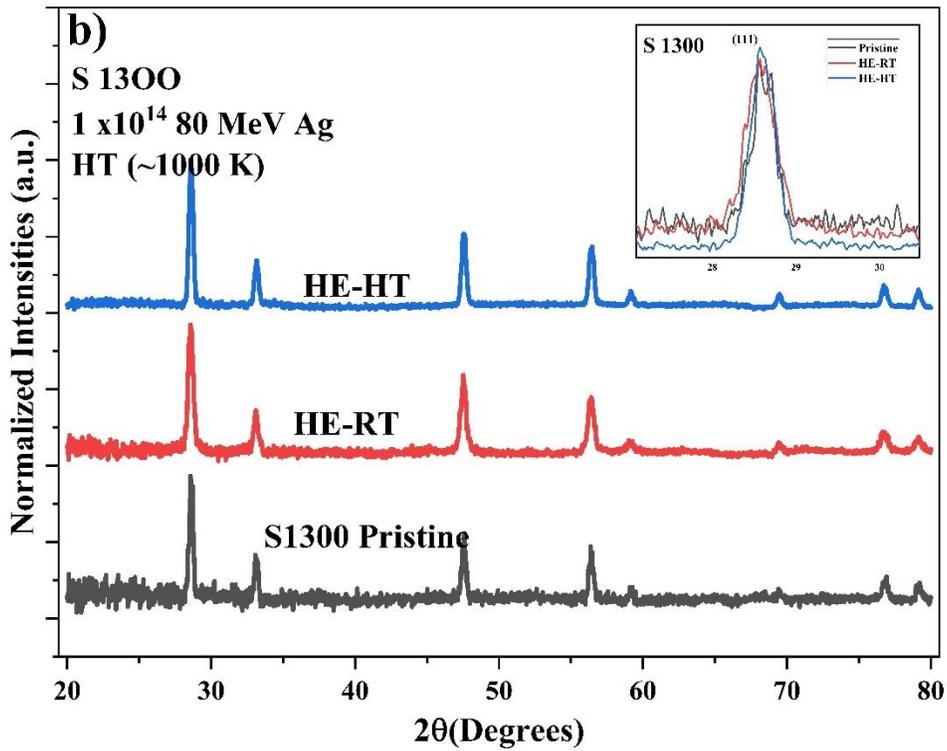

Figure No. 4. GIXRD pattern of 5-mol% Gd-doped Ceria. a) S800, b) S1300

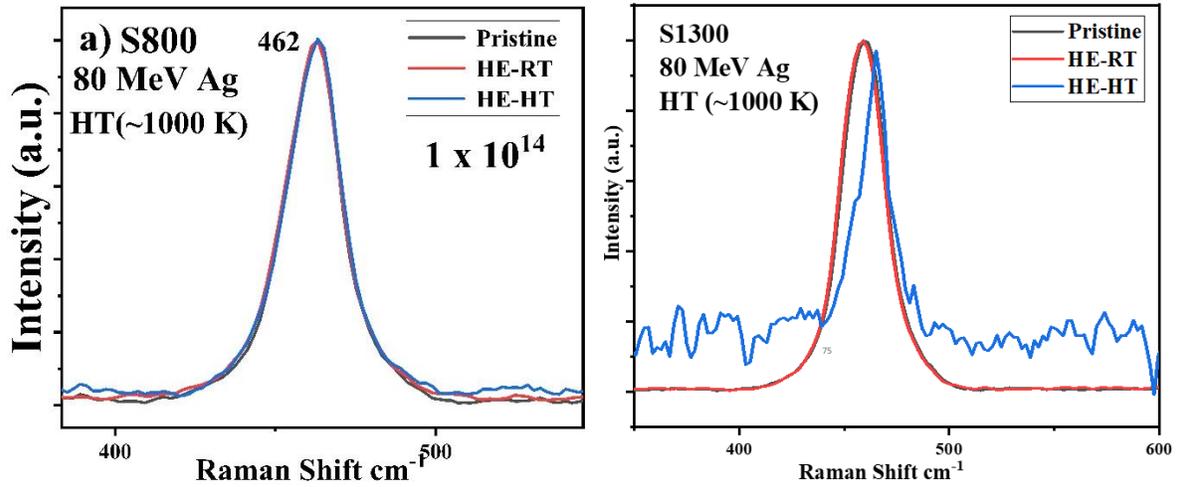

Figure No. 5. Raman spectra of (a) S800 irradiated at the indicated fluences (ions/cm$^2$), (b) S1300 irradiated at the indicated fluences (ions/cm$^2$).

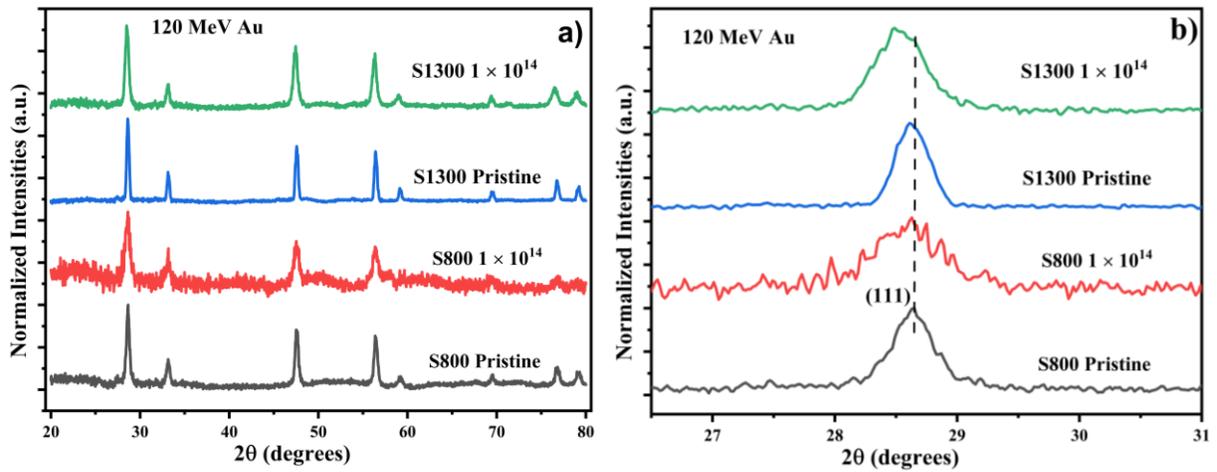

**Figure 6.** (a) GIXRD pattern of irradiated samples at the indicated fluences (ions/cm$^2$); (b) Zoomed view of (111) diffraction maxima of samples.

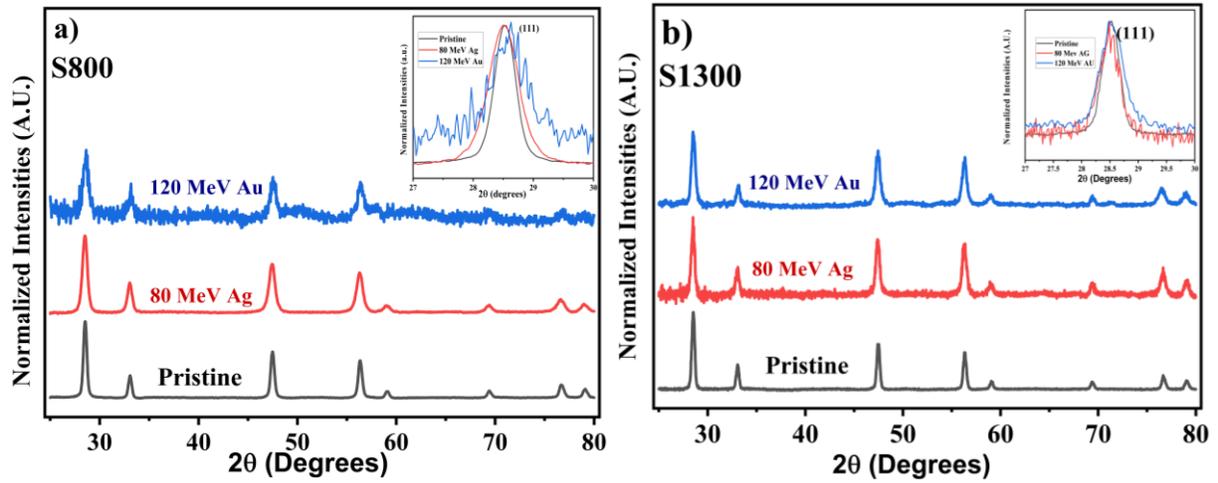

**Figure 7.** (a) GIXRD pattern of irradiated S800 samples at the indicated fluences (ions/cm$^2$); (b) GIXRD pattern of irradiated S1300 samples at the indicated fluences (ions/cm$^2$); Insets Zoomed view of (111) diffraction maxima of samples.

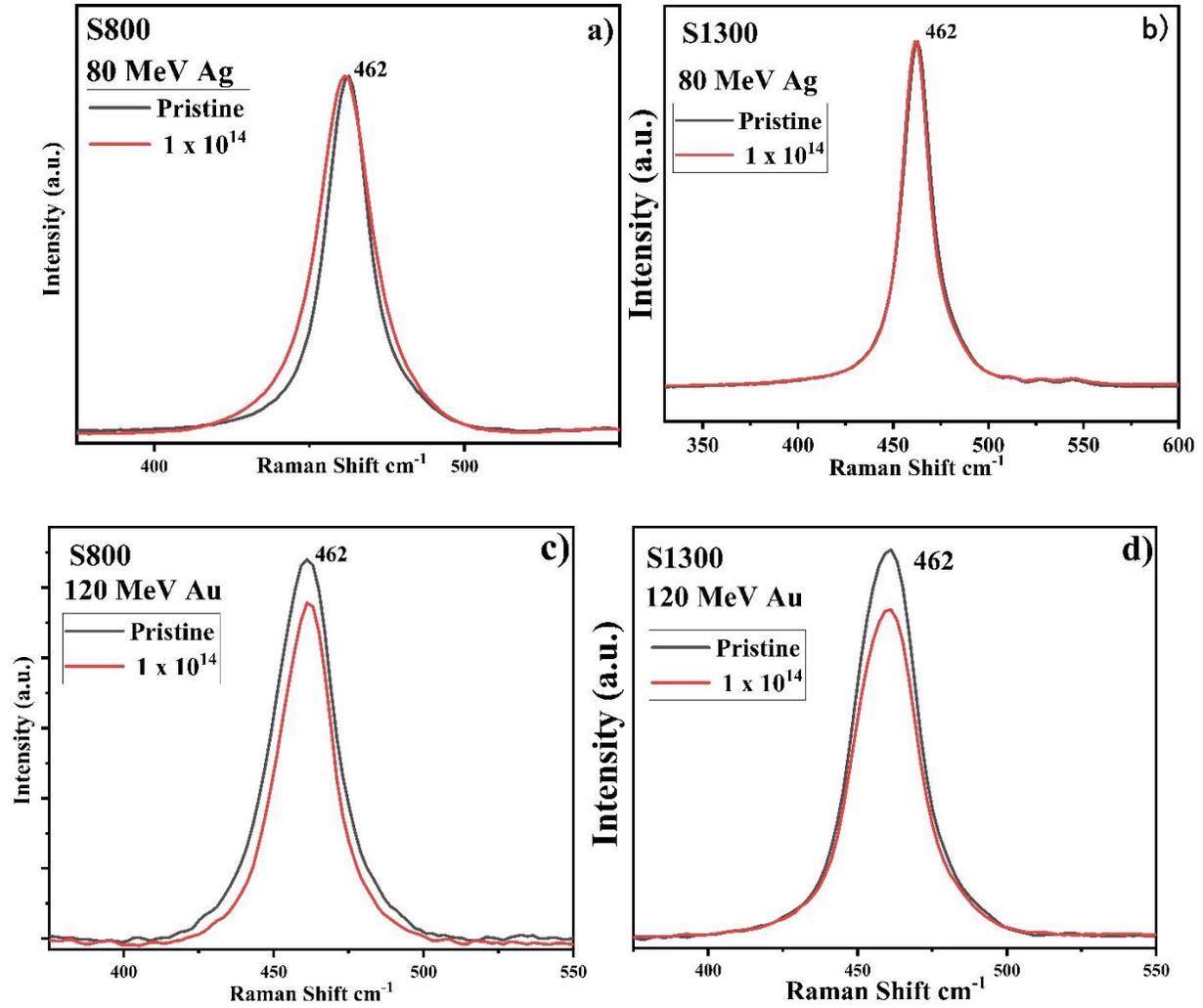

Figure 8. Raman spectroscopy spectra of electronic energy loss dependency, a). S800 irradiated with 80MeV Ag, b). S1300 irradiated with 80 MeV Ag, c). S800 irradiated with 120MeV Ag and d). S1300 irradiated with 120 Mev Ag.

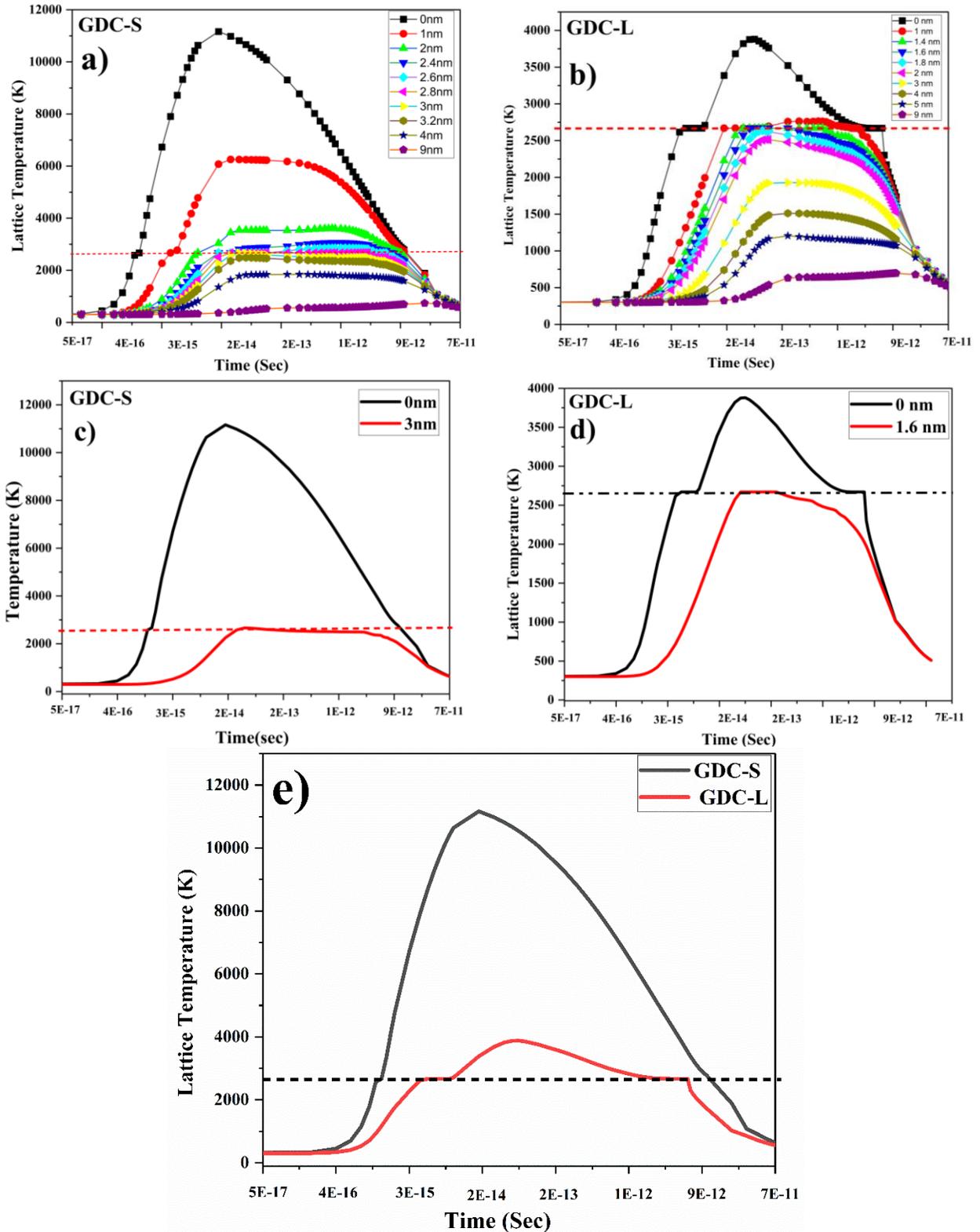

**Figure 9:** Inelastic Thermal Spike simulation results for GDC-S and GDC-L irradiated with 80 MeV Ag. The dashed lines indicate the material's melting point. (a) Variation in lattice temperature with time win GDC-S where the estimated track radius is ~ 3 nm, (b) Variation in lattice temperature with time in GDC-L where the estimated track radius is ~ 1.6 nm, (c) Comparison of lattice temperature in 0 nm and 3 nm in GDC-S, (d) Comparison of temperature in 0 nm and 1.6 nm in GDC-L, (e) Comparison between the maximum lattice temperature in GDCS- and GDC-L.

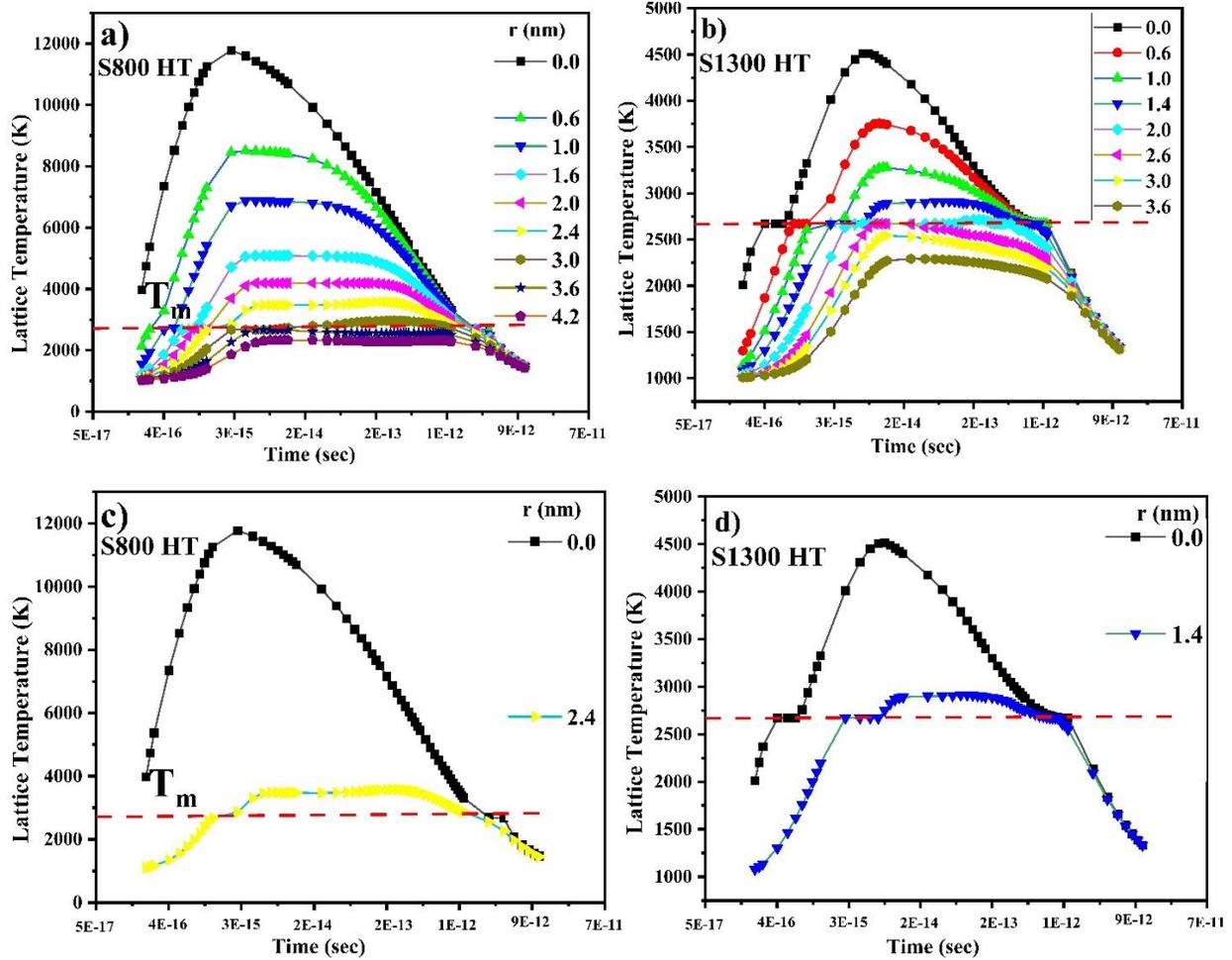

Figure No. 10. Inelastic Thermal Spike simulation results for S800 and S1300 irradiated with 80 MeV Ag at room temperature and high temperature (~1000 K). The dashed lines indicate the material's melting point. (a) Variation in lattice temperature with time win S800 at HT where the estimated track radius is ~ 2.4 nm, (b) Variation in lattice temperature with time in S1300 at HT where the estimated track radius is ~ 1.4 nm, (c) Comparison of lattice temperature in 0 nm and 2.4 nm in S800, (d) Comparison of temperature in 0 nm and 1.4 nm in S1300.

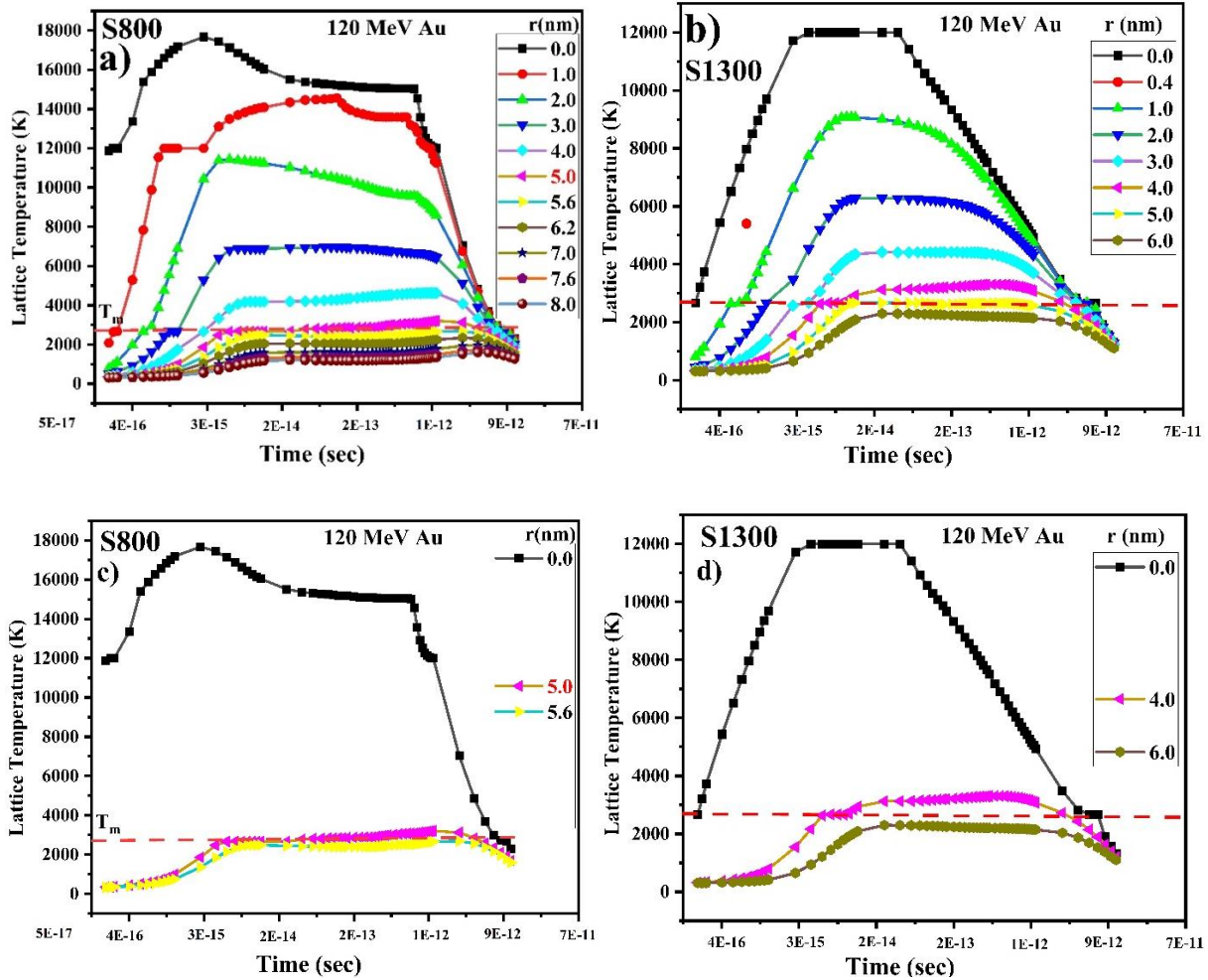

Figure No. 11. Inelastic Thermal Spike simulation results for S800 and S1300 irradiated with 120 MeV Ag at room temperature. The dashed lines indicate the material's melting point. (a) Variation in lattice temperature with time win S800 where the estimated track radius is ~ 5.0 nm, (b) Variation in lattice temperature with time in S1300 where the estimated track radius is ~ 4.0 nm, (c) Comparison of lattice temperature in 0 nm and 5 nm in S800, (d) Comparison of temperature in 0 nm and 4 nm in S1300.

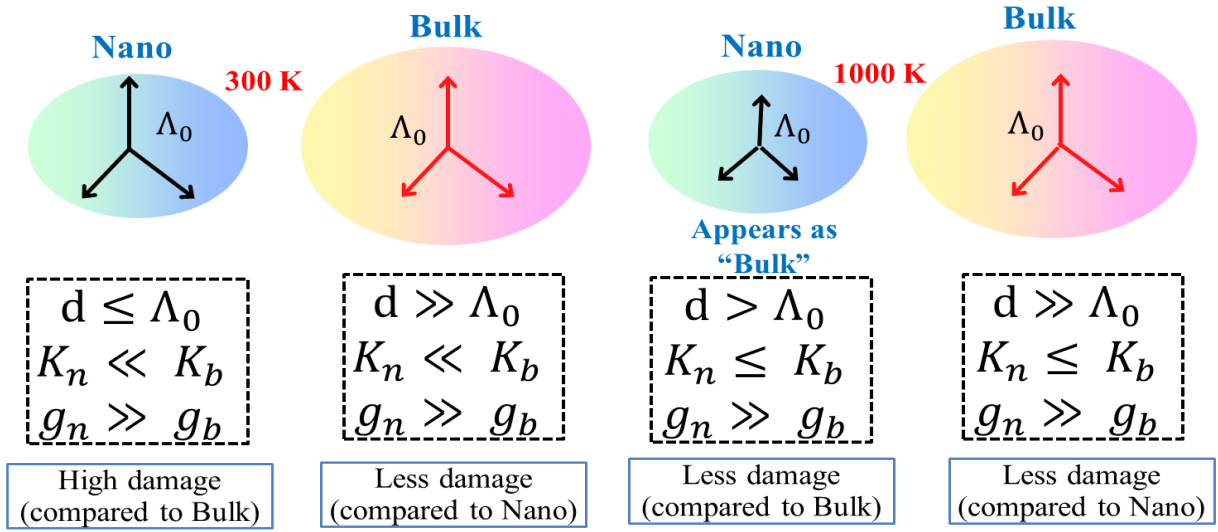

**Figure No. 11.** Schematics representation of in-elastic thermal spike model, where mainly phonon mean free path variation is restricted at high temperature.